\documentclass[conference]{IEEEtran}
\IEEEoverridecommandlockouts

\usepackage{algorithm}
\usepackage{algpseudocode}

\usepackage[style=ieee]{biblatex}
\bibliography{references}

\title{\LARGE \bf
Prebunking Design as a Defense Mechanism Against Misinformation Propagation on Social Networks
}

\newcommand{\Graph}{G}
\newcommand{\Verts}{V}
\newcommand{\Edges}{E}

\newcommand{\InNeigh}{N_\textrm{in}}

\newcommand{\graphprobs}{p}
\newcommand{\NN}{\mathbb{N}}
\newcommand{\RR}{\mathbb{R}}

\newcommand{\prob}{\mathbb{P}}
\newcommand{\expec}{\mathbb{E}}
\newcommand{\ra}{\rightarrow}
\newcommand{\centernode}{c}
\newcommand{\pdis}{\mathcal{P}}

\usepackage{amsmath,amssymb,amsfonts}
\usepackage{amsthm}
\usepackage{graphicx}
\usepackage{textcomp}
\usepackage{xcolor}
\usepackage{subcaption}

\newtheorem{assumption}{Assumption}
\newtheorem{theorem}{Theorem}
\newtheorem{lemma}[theorem]{Lemma}
\newtheorem{corollary}[theorem]{Corollary}

\begin{document}

\title{Prebunking Design as a Defense Mechanism Against Misinformation Propagation on Social Networks}

\author{\IEEEauthorblockN{1\textsuperscript{st} Yigit Ege Bayiz}
\IEEEauthorblockA{\textit{Electrical and Computer Engineering} \\
\textit{The University of Texas at Austin}\\
Austin, Texas, USA \\
egebayiz@utexas.edu}
\and
\IEEEauthorblockN{2\textsuperscript{nd} Ufuk Topcu}
\IEEEauthorblockA{\textit{Aerospace Engineering and Engineering Mechanics} \\
\textit{The University of Texas at Austin}\\
Austin, Texas, USA \\
utopcu@utexas.edu}
}

\maketitle
\thispagestyle{empty}
\pagestyle{empty}

\begin{abstract}
The growing reliance on social media for news consumption necessitates effective countermeasures to mitigate the rapid spread of misinformation. Prebunking, a proactive method that arms users with accurate information before they come across false content, has garnered support from journalism and psychology experts. We formalize the problem of optimal prebunking as optimizing the timing of delivering accurate information, ensuring users encounter it before receiving misinformation while minimizing the disruption to user experience. Utilizing a susceptible-infected epidemiological process to model the propagation of misinformation, we frame optimal prebunking as a policy synthesis problem with safety constraints. We then propose a policy that approximates the optimal solution to a relaxed problem. The experiments show that this policy cuts the user experience cost of repeated information delivery in half, compared to delivering accurate information immediately after identifying a misinformation propagation.
\end{abstract}

\section{INTRODUCTION}
Social media have become an integral part of modern communication, with more than $70\%$ of adults in the U.S. using at least one social media service \cite{PewResearch-2021-FactSheet}. At least $60\%$ of these users use \textit{social networking platforms}, a type of social media platform in which users build social networks and communicate with other users who share similar interests, opinions, or backgrounds. These platforms allow their users to easily share information and their opinions.

However, the ease of information sharing on social networking platforms has also led to the rise of \textit{misinformation} and fake news. Misinformation can spread faster and more broadly than accurate information, causing significant harm to public discourse. The consequences of misinformation can range from confusion to substantial damage to public health \cite{swire2020public}, or political manipulation \cite{benkler2018network}. The widespread adoption of social networking platforms and the growing reliance on social networking services as sources of news consumption exacerbates this issue, often resulting in significant portions of the population believing in unsubstantiated or provably incorrect claims, making it difficult to correct these false beliefs later through debunking.

\textit{Prebunking} has emerged as a promising solution for mitigating the spread of misinformation. Unlike debunking, which aims to correct false information after its spread, prebunking exposes people to factual information before they encounter false claims. The idea is to inoculate the public against misinformation, akin to a vaccine \cite{vanDerLinden-2017-Prebunking}. Studies indicate that this preemptive approach has the potential to reduce the spread of misinformation \cite{Lendowsky-2017-BeyondMisinfo}, without relying on media censorship.

The effectiveness of prebunking depends on the timing of delivery of the factual information. It is clear that for prebunking to work as intended the user has to see the factual information before seeing the related misinformation. Furthermore, the effectiveness of prebunking depends on the time gap between the delivery of the factual information and the arrival of the misinformation. Research in characterizing this relation has been sparse, leaving a significant gap in understanding the optimal timing for successful prebunking interventions \cite{pennycook2021psychology}. Another consideration is the effect of prebunking on the user-experience. Too frequent delivery of factual information can push users to other social networking platforms that do not utilize prebunking, rendering prebunking ineffective. 

We formalize optimal prebunking as a mathematical optimization problem over the timing of factual information delivery. The primary objective is to ensure that users encounter the factual information before being exposed to corresponding misinformation while minimizing any disruptions to the user-experience. We model the effect of factual information on user-experience by adopting a \textit{leaky bucket} cost, that is an exponentially decaying cumulative count over the number of prebunking related factual informations the user receives. Minimizing this cost discourages repeated and frequent delivery of factual information, minimizing the impact of prebunking on user-experience.

We use the susceptible-infected (SI) epidemiological model to characterize misinformation propagation and frame optimal prebunking as a policy synthesis problem with constraints to ensure the user receives the factual information before the corresponding misinformation. Building on this framework, we propose a locally optimal policy that attempts to minimize the leaky bucket cost, while ensuring that the user receives factual information prior to misinformation that is propagating over the social network.

\textbf{Contributions}
\begin{itemize}
    \item We present the problem of optimally delivering pre-generated factual information for the purpose of prebunking as a policy synthesis problem.
    \item We analyze two baseline methods that provide guaranteed delivery of factual information before its corresponding misinformation.
    \item We provide a third method that approximates the optimal solution to a relaxed problem and provides the same guarantees as the previously mentioned methods with less cost.
\end{itemize}

\section{RELATED WORKS}
\subsection{Countering Misinformation}
We classify misinformation countering methods into four categories, \textit{censorship}, \textit{debunking}, \textit{prebunking}, and \textit{identification}, The first three categories all attempt to reduce the impact of misinformation. Censorship refers to any method, which aims to curb misinformation spread by attempting to control the propagation of information in the network \cite{LiuBuss2020Censor, bayiz2022countering}. Censorship is common in social networking platforms, yet it raises significant issues relating to the freedom of speech. 

Debunking refers to correcting misinformation by providing users with correct information after the misinformation has already spread, whereas prebunking refers to issuing correct information before misinformation propagates. An automated example of debunking is the numerous automated fact-checking methods that all aim to debunk misinformative text content \cite{Guo2022fatchecking, augenstein-etal-2019-multifc}. The current understanding of social psychology indicates prebunking to be superior to debunking in terms of its effectiveness in countering misinformation \cite{vanDerLinden-2017-Prebunking, pennycook2021psychology, Ecker2022}. In this paper, we contribute to the automation of prebunking by developing algorithms for automatically optimizing the delivery times of prebunks to the users in a social networking platform. 

Identification refers to any method that aims to detect misinformative content within a social network. These models often utilize natural language processing models \cite{CHATURVEDI201865, mohtarami-etal-2018-automatic}. Del Vicario et al. \cite{vicario2016spreading} has shown that the propagation characteristics of misinformation admit detection without relying on content classification. More recently, Shaar et al. \cite{shaar-etal-2020-known} introduce a method with which to identify already fact-checked claims. In this paper we do not use misinformation detection directly. However, we assume we already know the misinformation content, thus accurate misinformation detection remains a pre-requisite for the methods we present.

\subsection{Rumor Propagation Models}
Determining optimal times for prebunking deliveries requires accurate estimations for when the misinformation will arrive to the user of interest. This estimation requires a rumor propagation model. In this paper, we rely heavily on \textit{epidemic models} \cite{hill2010infectious}, also known as compartmental models, to model misinformation propagation. As their name suggests, these models are based on epidemiology, and model rumor propagation by partitioning users into different categories, such as \textit{susceptible}, or \textit{infected}, and then define rules by which these partitions interact over time. There is a wide range of epidemic models that are used in misinformation modeling \cite{raponi2022fake}. The most among these are,  SI \cite{Krishnasamy2014SI}, SIR \cite{zhao2013sir, wang2017sir}, SIS \cite{kimura2009efficient, jin2013epidemiological} models. SI (susceptible-infected) models are easy to model and often are the only models that permit analysis with arbitrary graph models. SIR (susceptible-infected-recovered) and SIS (susceptible-infected-susceptible) refine the SI model, making them more accurate without introducing significant computational complexity to simulations. Despite these refinements, SI propagation still finds use due to its simplicity, and due to having behavior that is comparable to SIS and SIR models for the initial phase of misinformation propagation, which is the most critical phase for countering misinformation. Throughout this paper, we use an SI model to estimate misinformation propagation.
\section{PRELIMINARIES}
\subsection{Discrete-Time SI Model on Graphs}
A discrete-time \emph{susceptible-infected} (SI) model \cite{hill2010infectious} is a widely-utilized framework for understanding epidemic spread through a network. It models how a contagion spreads in a network over consecutive iterations until it cannot infect any more nodes in the network.

An SI propagation on a directed graph $\Graph = (\Verts, \Edges, [p_{ij}])$ defines how an \textit{infected subset} $I_t$ of the entire population $\Verts$ evolves over discrete time steps $t$. At each time $t$, The SI model defines a probability distribution on $I_{t+1}$, depending on $I_{t}$, which is as follows,
\begin{equation}\label{dynamics_inf}
    \prob(j \in I_{t+1} | j \not\in I_t, I_t) = 1 - {\sum_{\substack{i \in I_t \\ i \neq j} }{(1 - p_{ij})}}.
\end{equation}
Intuitively, at each time step, each infected node $i \in I_t$ spreads its infection to all susceptible nodes $j \in S_t$, with probability $p_{ij}$. Each node that gets infected remains infected forever. The simulation proceeds until no new nodes can become infected.

\subsection{Modelling Misinformation Propagation}
A \textit{social networking platform} is an online social media platform that allows its users to communicate, share opinions, and form social networks. To date, popular examples of such services include Facebook, LinkedIn, Instagram, and Twitter. These platforms provide a user interface for users to communicate with each other by either broadcasting messages, commenting on other's messages, or sending direct messages to other users. In addition, these platforms can display advertisements, or notifications to the users, depending on the purpose or the design of the specific social networking platform. 

We let the term \textit{content} refer to any message or information that the users can interact with on a social networking platform. Contents can either be \textit{static} or \textit{contagious}. Static contents is those that the social networking platform provides to the users. These contents do not propagate between users. Examples of such static contents are advertisements, notifications, and alerts that the social networking platform shows to its users. Contagious content, on the other hand, spreads between users. In a social networking platform, contagious content can be user posts, messages, comments, or replies. 

We define a \textit{contagion} as a collection of contagious contents conveying the same claim. Contagions represent the rumors that spread from user to user on a social networking platform. For example, all contagious content claiming that ``5G causes COVID-19'' constitutes a single contagion. 

A \textit{misinformation} denotes a contagion we do not want to spread in the network. Often, misinformations contain information that is incorrect, or harmful to the individual. We denote the set of all misinformations that exist on the social networking platform by $M = \{M_0, M_1, \dots \}$, where we assume $M$ to be a countable set, either finite or infinite. Referring to misinformation as a countable noun as we have done in this paper is clearly a diversion from existing literature. However, as we are using the term to refer to a specific contagion, and since there are many different contagions that can be classified as misinformation, we will use the plural noun misinformations when referring to more than one misinformation. 

We model the social networking platform as a weighted directed graph $\Graph = (\Verts, \Edges, \graphprobs)$ where the set of vertices $\Verts$ represents users, the set of edges $\Edges$ represents possible connections between users, and $\graphprobs_{ij}$ are normalized edge weights, representing the probability of misinformation propagation over the associated edge $ij \in \Edges$. For all $ij\not \in \Edges$ we define $p_{ij} = 0$.

We define the local network around a central user $\centernode$ as the subgraph $\Graph_\centernode = (\Verts_\centernode, \Edges_\centernode, \graphprobs)$ of $\Graph$ denoting the $m$-neighborhood of the $\centernode$. That is, $\Graph_\centernode$ is the subgraph of $\Graph$ containing all users that have a shortest path length of at most $m$ to $\centernode$. Here, $m$ is a spatial horizon parameter we can choose as any integer between $1$ to $\textrm{diam}(\Graph)$, with higher values providing a more accurate model, with a larger number or users. We also assume without generality that $\Verts_\centernode = \{1,2,\dots, N\}$, meaning we have $N$ users in the local network and they are indexed from $1$ to $N$.

We model the misinformation propagation in the local neighborhood of the user $\centernode$ as a susceptible-infected (SI) process on the local subgraph $\Graph_\centernode$, where $\graphprobs$ is the infection spreading probability. That is, for misinformation $M_k$, and time $t \in \NN$, $I_t^k \subseteq \Verts_\centernode$ denotes the set of users that are actively spreading the misinformation, and $S_t^k = \Verts\setminus I_t^k$ denote the rest of the users.  Assumption \ref{assumption:model} collects all of these effects into a formal model that describes the propagation around user $\centernode$.

\begin{assumption}\label{assumption:model}
    The propagation dynamics of each misinformation $M_k$ is independent from the other misinformations. In particular for all $k \in \NN$,
    \begin{equation}
    \begin{split}
        \prob(j \in I_{t+1}^{k} &| j \in S_t^{k}, \, I_{t}^{k}\neq \emptyset, \, I_{t}^{k})\\ &= 1 - \prod_{i \in \InNeigh(j)\cap I_{t}^{k}} (1 - \graphprobs_{ij})
    \end{split}
    \end{equation}
\end{assumption}

In addition to the SI dynamics, each user $i$ has a small probability $q^k_i(t)$ representing the probability of the misinformation $M_k$ originating on any node $i$ of the network. This additional probability captures the propagation effects that are unaccounted for in our model, such as misinformation propagation outside the local neighborhood of user $\centernode$, or outside the social networking platform.

In practice, it is difficult to estimate the probabilities $q^k_i(t)$ as users can differ vastly in their tendencies to generate misinformation. Furthermore, under the SI model, each misinformation $M_k$ cannot infect a user more than once. Thus it is impossible to obtain any statistical information on $q^k_i(t)$ before the node $i$ is gets infected, rendering both Bayesian and frequentist estimations of $q^k_i(t)$ infeasible. Thus, throughout this paper, we stipulate that $q^k_i(t)$ are equal to a single probability $q_i(t)$. This probability $q_i(t)$ can be thought as the average of $q^k_i(t)$ across all $k$. Unlike $q^k_i(t)$, it is possible to learn $q_i(t)$ by employing a frequentist estimation over all misinformation cascades.

At each time step we allow any user to generate only one misinformation. Also, as we model the propagation of misinformation solely within the local network we assume each user can only generate a new misinformation that is not already propagating in the network. For a more rigorous model of misinformation generation, assume an ordering on $M$ such that at each time $t$, $i$'th user can only generate misinformation $M_{Nt + i}$. Thus, at time $T$, for any $i\in V_c$ we always have $I_t^{NT + i} = \emptyset$ for all $t < T$. Furthermore, if $I_T^{NT + i} = \emptyset$ then we have $I_t^{NT + i} = \emptyset$ for all $t > T$, meaning if user $i$ misses to generate misinformation $M_{NT + i}$ at time $T$, they can never generate it again, and that misinformation will simply be a formal object that does not exist in the real-world network. The non-existance of $M_{NT + i}$ is not an issue however, as the misinformation set $M$ is purely a formal construct that allows us to formulate the prebunking problem rigorously. The misinformations $M_k$ in this regard, do not correspond to misinformations in the real world network per se, but potential misinformations that can exist. Whether they do exist or not, and exactly which contents they contain only becomes apparent after they start propagating in the social networking platform. Assumption \ref{assumption:misinfoinit} provides the formal definition for how new misinformations originate.

\begin{assumption}\label{assumption:misinfoinit}
    For all times $t\in \RR$ and for all users $i \in V_c = \{1,2,\dots,N\}$ we have
    \begin{align}
            I_t^{Nt + i} = \begin{cases}
                \{i\}, \quad &\textrm{with prob.} \quad q_i(t) \\
                \emptyset, \quad &\textrm{with prob.} \quad 1 - q_i(t).
            \end{cases}
    \end{align}    
\end{assumption}

\section{OPTIMAL PREBUNKING PROBLEM}
\begin{figure*}
\begin{subfigure}{.2\textwidth}
  \centering
  \includegraphics[width=.9\linewidth]{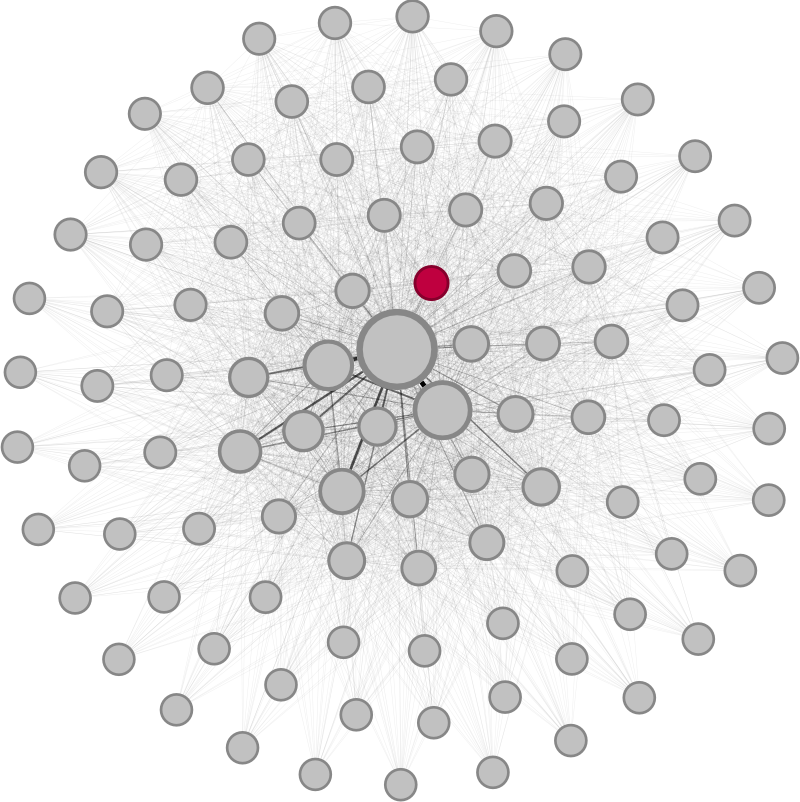}
  \caption{time = 0}
  \label{fig:smallnobunkT0}
\end{subfigure}%
\begin{subfigure}{.2\textwidth}
  \centering
  \includegraphics[width=.9\linewidth]{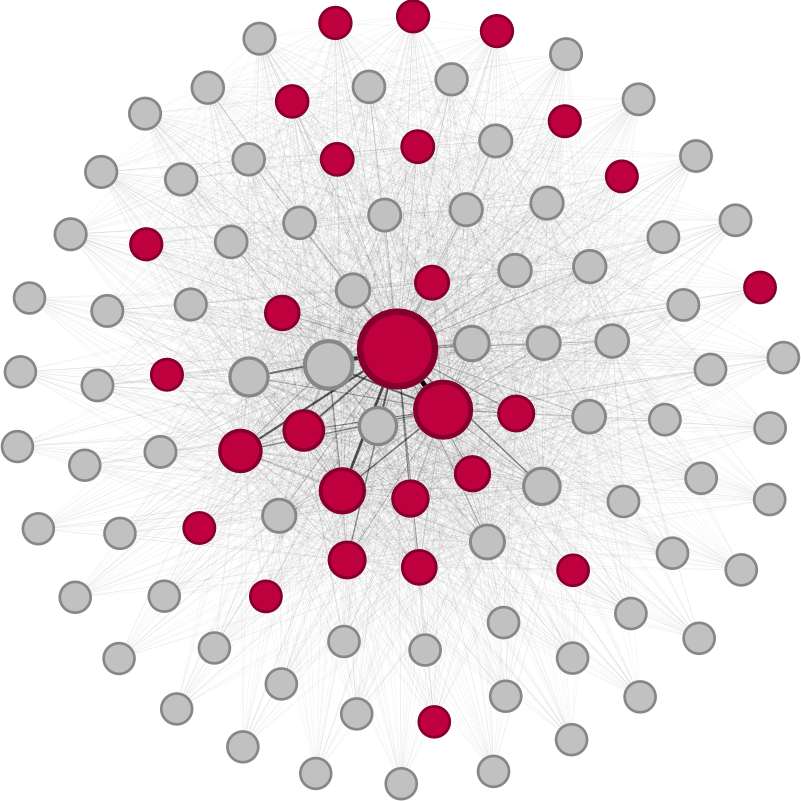}
  \caption{time = 5}
  \label{fig:smallnobunkT5}
\end{subfigure}%
\begin{subfigure}{.2\textwidth}
  \centering
  \includegraphics[width=.9\linewidth]{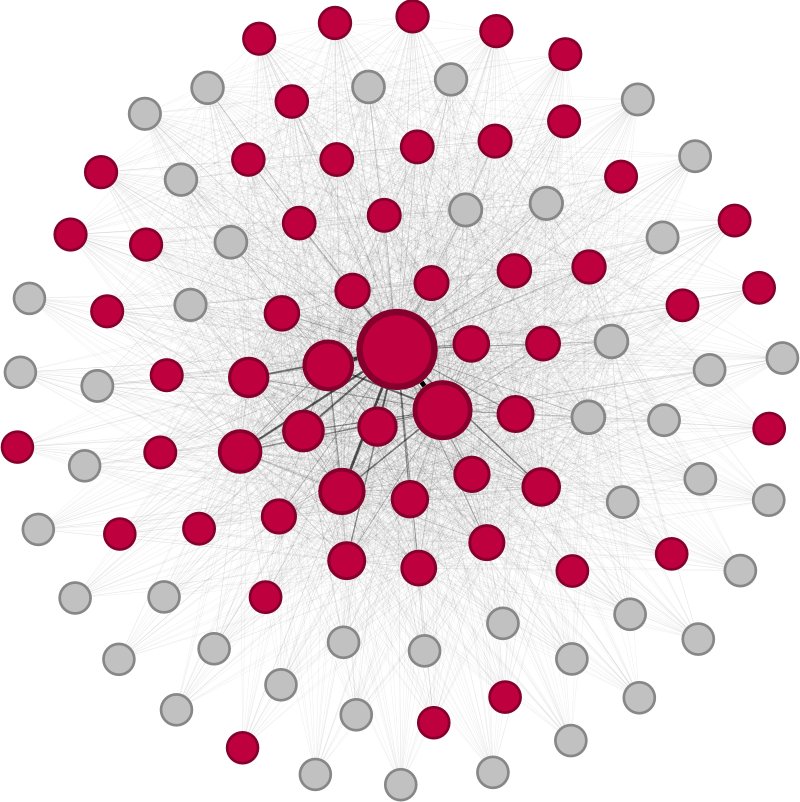}
  \caption{time = 10}
  \label{fig:smallnobunkT10}
\end{subfigure}%
\begin{subfigure}{.2\textwidth}
  \centering
  \includegraphics[width=.9\linewidth]{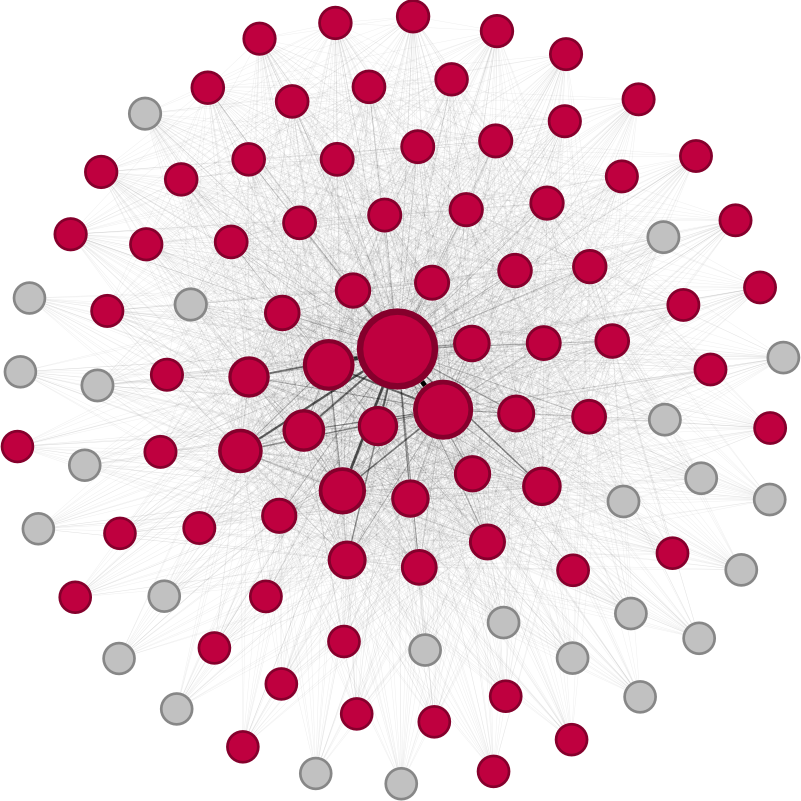}
  \caption{time = 15}
  \label{fig:smallnobunkT15}
\end{subfigure}%
\begin{subfigure}{.2\textwidth}
  \centering
  \includegraphics[width=.9\linewidth]{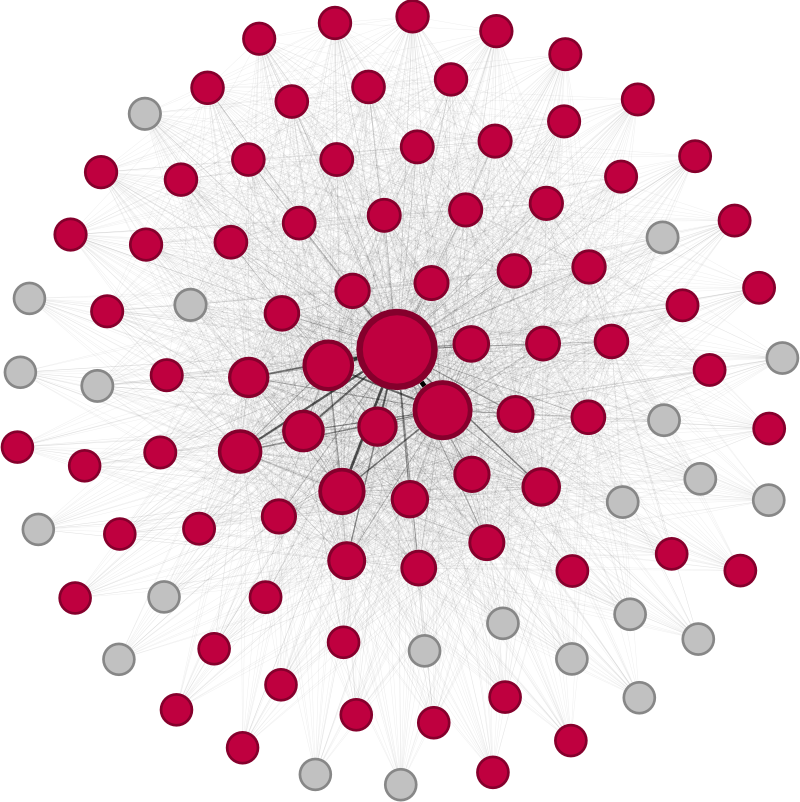}
  \caption{time = 20}
  \label{fig:smallnobunkT20} 
  \end{subfigure}%
\\ 

  \begin{subfigure}{.2\textwidth}
  \centering
  \includegraphics[width=.9\linewidth]{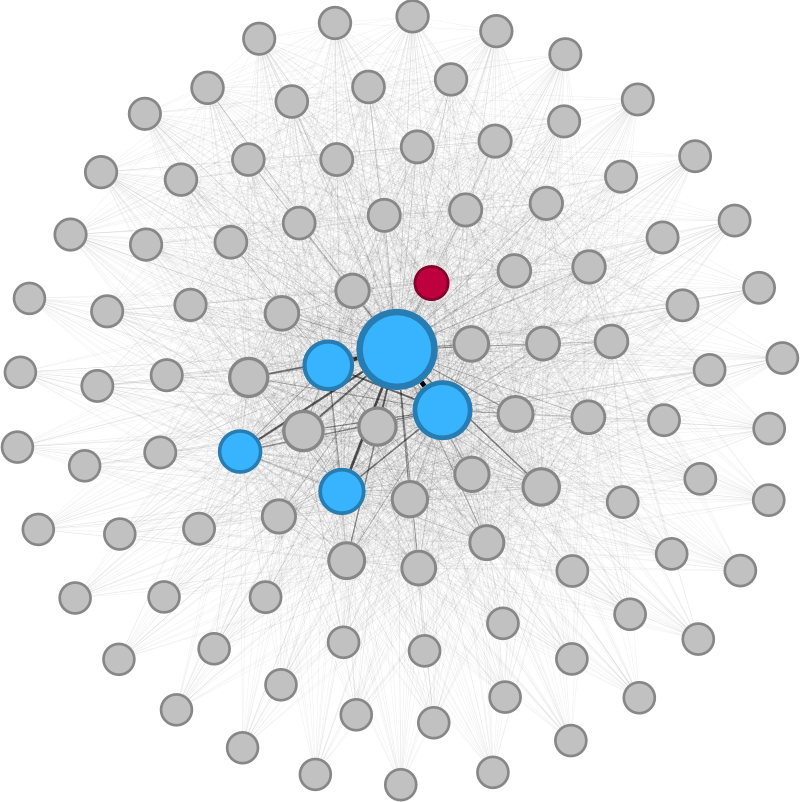}
  \caption{time = 0}
  \label{fig:smallprebunkT0}
\end{subfigure}%
\begin{subfigure}{.2\textwidth}
  \centering
  \includegraphics[width=.9\linewidth]{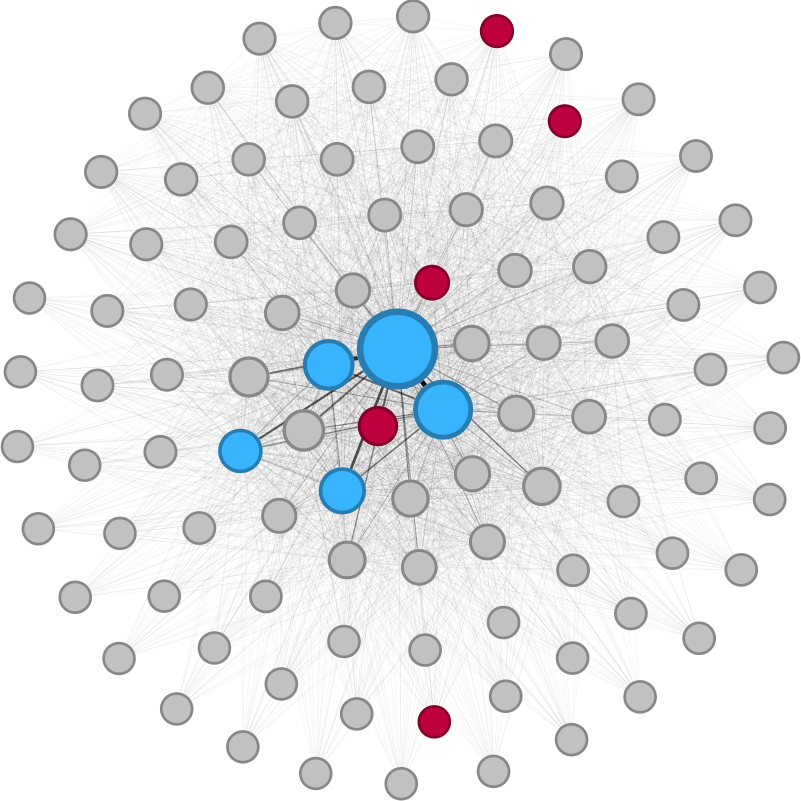}
  \caption{time = 5}
  \label{fig:smallprebunkT5}
\end{subfigure}%
\begin{subfigure}{.2\textwidth}
  \centering
  \includegraphics[width=.9\linewidth]{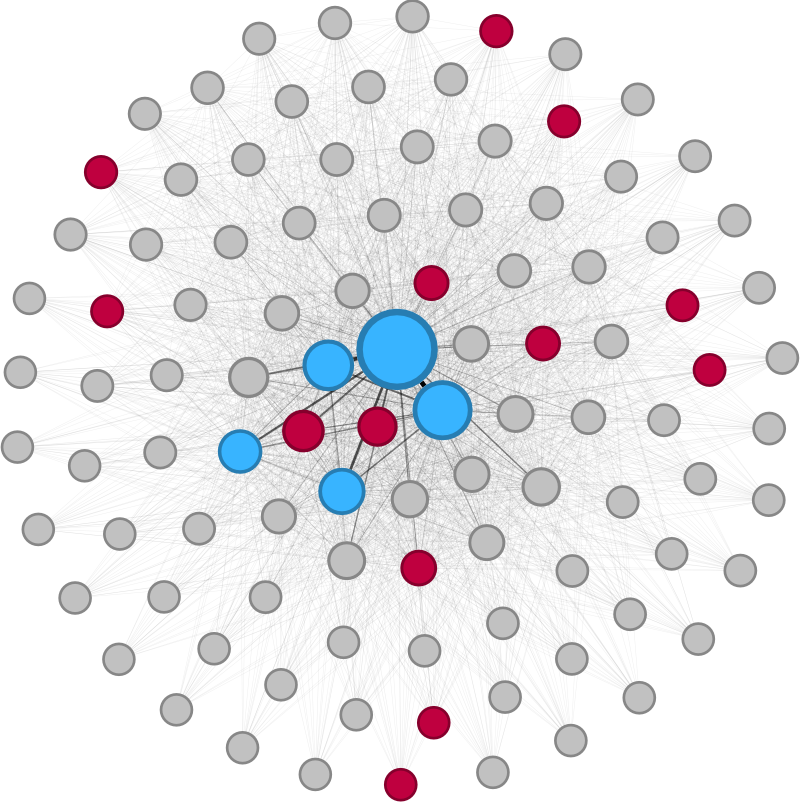}
  \caption{time = 10}
  \label{fig:smallprebunkT10}
\end{subfigure}%
\begin{subfigure}{.2\textwidth}
  \centering
  \includegraphics[width=.9\linewidth]{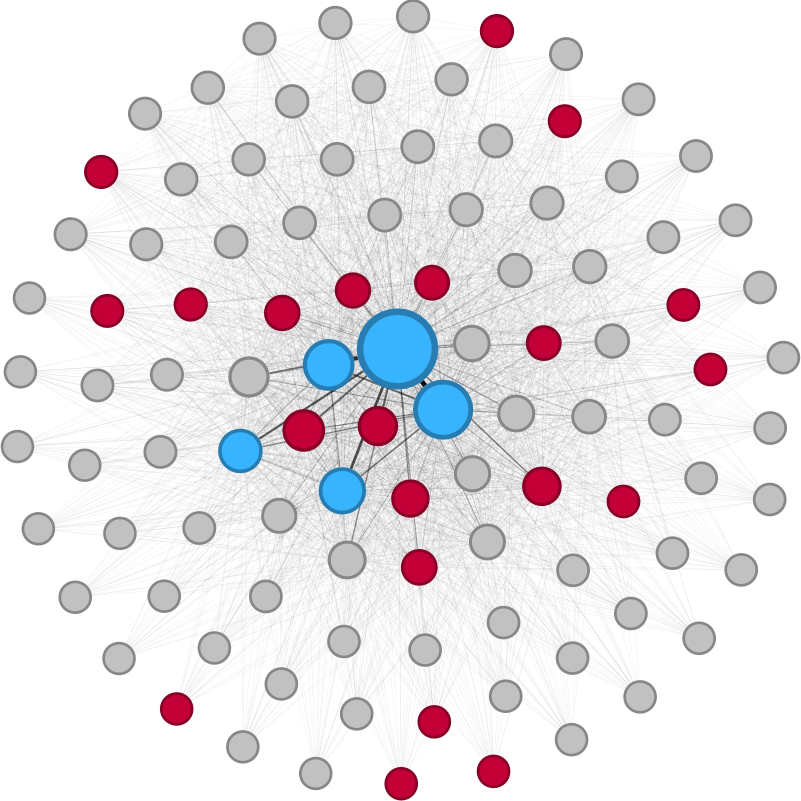}
  \caption{time = 15}
  \label{fig:smallprebunkT15}
\end{subfigure}%
\begin{subfigure}{.2\textwidth}
  \centering
  \includegraphics[width=.9\linewidth]{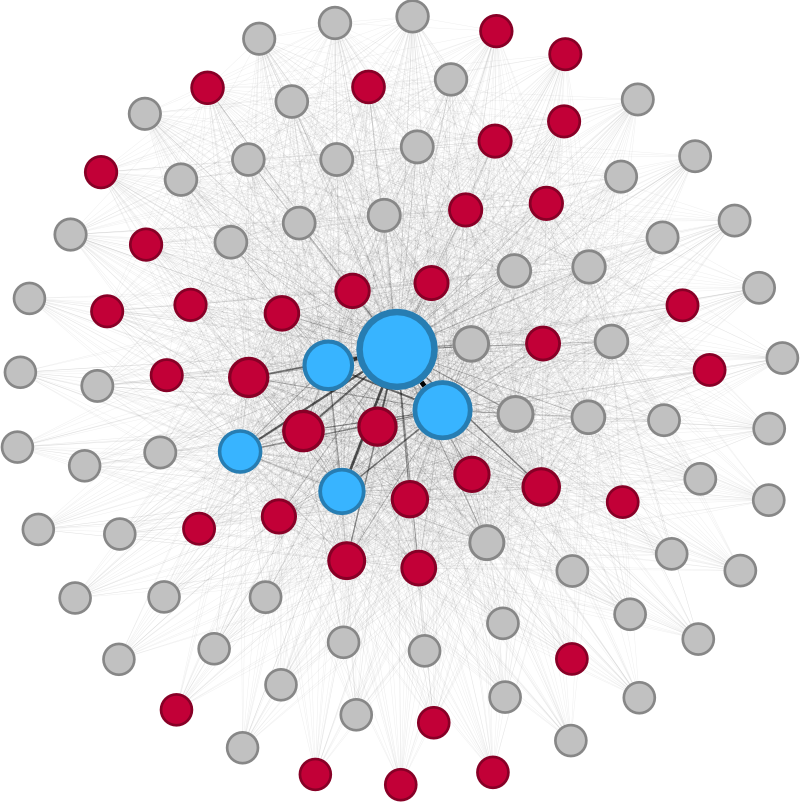}
  \caption{time = 20}
  \label{fig:smallprebunkT20}
\end{subfigure}%
\caption{Two simulated misinformation propagations in a small network of users: Subfigures (a-e) show propagation with no prebunking, and subfigures (f-j) show propagation with prebunking. Dark red vertices represent misinformed users $I_t$, light blue vertices represent inoculated users, gray vertices represent the rest of the users.}
\label{fig:small_network}
\end{figure*}

To guard user $\centernode$ against misinformation, we employ \textit{prebunking}. Prebunking relies on inoculating the users in the network by providing the users with correct information before they encounter a misinformation that contradicts the correct information. The idea of prebunking is similar to that of vaccination, where an early intervention to misinformation can immunize some users against it, reducing misinformation spread in the population. Fig. \ref{fig:small_network} shows the effect of prebunking in a small network of users.

We define a \textit{prebunk} as a piece of static content that the social networking platform generates and delivers to the user before the user receives the associated misinformation. That is, we associate a prebunk $P_k$ with each misinformation $M_k$ and impose the condition that the central user $\centernode$ must always receive $P_k$ before $M_k$. Since the social networking platform artificially generates each prebunk $P_k$, we assume that they are static, and do not follow an SI propagation and that their delivery is under full control of the social networking platform. At each time $t$, the social networking platform can pick any number of prebunks $P_k$ to deliver to user $\centernode$. We let $A_t \subseteq {P}$, denote the the subset of prebunks the social networking platform delivers to user $\centernode$ at time $t$, where $P = \{P_1\dots P_k\}$.

We refer to the rule with which the social networking platform determines the subset $A_t$ of prebunks, as a \textit{randomized stationary Markov policy}, or simply policy. Formally, the policy is a function that maps the current set of infected nodes $(I_t^1 \dots I_t^k)$ to some probability distribution within the set $\pdis(2^P)$ of all probability distributions over the power set of prebunks $2^P$. That is, $\pi: \bigotimes_{k=1}^{K} 2^V \ra \pdis(2^P)$. Intuitively, $\pi(I_t^1 \dots I_t^k)$ assigns a probability to all possible prebunk subsets the social network platform can deliver to user $\centernode$, based on where the sets $I^k_t$ of misinformation spreaders are located. The social network platform then randomly samples $\pi(I_t^1 \dots I_t^k)$ to determine $A_t \subseteq {P}$. We denote the probability of the social network platform choosing prebunks $a_t$ given misinformed sets $(I_t^1 \dots I_t^k)$ as $\prob_\pi(A_t = a_t| (I_t^1 \dots I_t^k))$. We call a policy deterministic if $\prob_\pi(A_t = a_t| (I_t^1 \dots I_t^k)) = 0$ for all possible $a_t$ except one. We also let the set of all policies be $\Pi$, and the set of all deterministic policies as $\Pi_{d}$. 

Finally, we define $X_k$ to be the random variable that denotes the time it takes for user $\centernode$ to encounter misinformation $k$. That is, 
\begin{equation}
    X_k = \min_{t\in\NN}\{\centernode \in I_t^k\}.
\end{equation}
Then, we can express the condition that prebunk $P_k$ will be delivered before misinformation $M_k$ as
\begin{equation}
    P_k \in \bigcup_{t = 0}^{X_k-1}{A_t}.
\end{equation}
Thus we write the optimal prebunking problem as the following policy synthesis problem,
\begin{subequations}\label{opt:min}
\begin{align}
    \min_{\pi \in \Pi} \quad & C(\pi)\\
    \text{s.t.} \quad & A_1,\dots, A_T \sim \pi(I_t^1 \dots I_t^k) \\
    &\forall k\in\NN, \quad \prob(P_k \in \bigcup_{t = 0}^{X_k-1}{A_t}) = 1.
\end{align}
\end{subequations}
Here, $C$ is a function over the policy space $\Pi$ reflecting the policy's negative impact on the user experience. Throughout our analysis, we define the cost $C$ as the maximum leaky bucket cost accumulated over a finite time horizon $T$. That is,
\begin{equation}
    C(\pi) = \max_{t \in [T]} \sum_{\tau=1}^t \beta^{t-\tau} \expec_\pi[|A_\tau|]. 
\end{equation}
Here, $\expec_\pi[|A_\tau|]$ denotes the expected number of prebunks the user receives at time $\tau$ under policy $\pi$; and $\beta$ is a forgetting parameter on the interval $(0,1]$, which describes how quickly the effect of prebunks decay over time. 

For ease of further reference we let
\begin{equation}
    c(\pi, t) =  \sum_{\tau=1}^t \beta^{t-\tau} \expec_\pi[|A_\tau|], \quad C(\pi) = \max_{t \in [T]} c(\pi, t).
\end{equation}
Also for any sequence $A = (A_1, \dots, A_T)$ of prebunk sets, we let 
\begin{equation}
    c(A, t) =  \sum_{\tau=1}^t \beta^{t-\tau} |A_\tau|, \quad C(A) = \max_{t \in [T]} c(A, t),
\end{equation}
so that 
\begin{equation}
    c(\pi, t) = \expec_{A\sim\pi}[c(A, t)], 
\end{equation}
and
\begin{equation}
    C(\pi) \geq \expec_{A\sim\pi}[C(A)],
\end{equation}
where the latter is due to Jensen's Inequality.

Thus, the original optimization problem \ref{opt:min} becomes the following constrained minimax problem,

\begin{subequations} \label{prob:optimization_minimax}
\begin{align}
    \min_{\pi \in \Pi}\max_{t \in [T]} \quad & c(\pi, t)\\
    \text{s.t.} \quad & A_1,\dots, A_T \sim \pi(I_t^1 \dots I_t^k), \\
    &\forall k\in\NN, \quad \prob(P_k \in \bigcup_{t = 0}^{X_k-1}{A_t}) = 1. \label{eq:opt:feasibility}
\end{align}
\end{subequations}

Lastly, we note that it is impossible to guard the center user $\centernode$ against a cascading misinformation $k$ if the misinformation starts from $\centernode$. As there is no way of estimating when a new misinformation cascade will start, there is no way to preemptively provide a prebunk to the central user $\centernode$ if a misinformation cascade starts on it. Thus, for theoretical analysis it is often necessary to assume that a misinformation cascade never starts at $\centernode$. We rephrase this assumption as follows:

\begin{assumption}\label{assumption:idealcenter}
    The central user $\centernode$ cannot start a misinformation cascade, and can only receive misinformation from its neighboring users. That is, $q_\centernode = 0$.
\end{assumption}

\section{DETERMINISTIC BASELINE POLICIES}
We start our analysis by introducing two baseline deterministic policies with feasibility guarantees. The first of these approaches is called \textit{early response}, described in Algorithm \ref{alg:pre}. In this policy, we provide the prebunks as soon as a new misinformation is spotted in the local neighborhood of the user. 

\begin{algorithm} 
    \caption{Early Response Policy}\label{alg:pre}
    \begin{algorithmic}
       \State For each $k$, observe misinformed nodes $I_t^k$ in the local neighborhood of the user.
       \State Choose action
       \begin{equation}
           A_t = \bigg\{P_k \in P\Big | I_t^k \not = \emptyset, \, P_k \not \in \bigcup_{\tau = 1}^{t - 1} A_\tau\bigg\}
       \end{equation}
    \end{algorithmic}
\end{algorithm}

It is clear that under Assumption \ref{assumption:idealcenter}, Algorithm \ref{alg:pre} ensures that the feasibility condition \ref{eq:opt:feasibility} holds, which we can show rigorously by observing that Assumption \ref{assumption:idealcenter} guarantees that the smallest $t$ for which $I_t \neq \emptyset$, satisfies $t < X_k$. Then by Algorithm \ref{alg:pre} we ensure
\begin{equation}
    P_k \in A_t \subseteq \bigcup_{\tau = 0}^{X_k-1}{A_\tau}.
\end{equation}

A weakness of Algorithm \ref{alg:pre} is that the timings of prebunk deliveries are purely determined by when the misinformations arrive. Thus if multiple different misinformations appear in rapid succession, the cost $C(A)$ can get large. For example, if misinformations $M_1,\dots,M_K$ all appear at time $t$, so that $I^k_t \neq \emptyset$ and $I^k_{t-1} = \emptyset$, for all $k \in [K]$, then under Algorithm \ref{alg:pre} we would have 
\begin{equation}
    A_t = \{P_1,\dots,P_K\},
\end{equation}
which implies that the accumulated cost $C(A)$ across all actions $A = (A_1,\dots, A_T)$ until the time horizon $T$ would satisfy
\begin{equation}
    C(A) = \max(c(A,t)) > c(A_t, t) > |A_t| = K.
\end{equation}
Thus, the accumulated cost over time is lower bounded by the number of simultaneous misinformation appearances. 

Given that the misinformation generation probabilities are constant across time, it is also possible to provide an exact closed-form expression for the policy cost of Algorithm \ref{alg:pre}, which can be seen in Lemma \ref{lem:pre_cost}.

\begin{lemma} \label{lem:pre_cost}
    Let $\pi \in \Pi_d$ denote the deterministic policy given by Algorithm \ref{alg:pre}. Suppose that for all $i \in V_c$, $q_i = q$ Then the instantaneous expected cost $c(\pi,t)$ is as follows,
    \begin{equation}
        c(\pi,t) = \frac{1 - \beta^t}{1 - \beta}\sum_{i \in V_\centernode} q_i.
    \end{equation}
    Furthermore,
    \begin{equation}
        C(\pi) = \frac{1 - \beta^T}{1 - \beta}\sum_{i \in V_\centernode} q_i.
    \end{equation}
    \begin{proof}
    The characterization of $c(\pi,t)$ follows due to
        \begin{align}
            c(\pi,t) &= \sum_{\tau=1}^t \beta^{t-\tau} \expec_\pi[|A_\tau|]\\ &= \sum_{\tau=1}^t \beta^{t-\tau} \sum_{i \in V_\centernode} q_i = \frac{1 - \beta^t}{1 - \beta}\sum_{i \in V_\centernode} q_i.
        \end{align} 
Since $c(\pi,t)$ is increasing in $t$ we have
    \begin{equation}
            C(\pi) = c(\pi,T) = \frac{1 - \beta^T}{1 - \beta}\sum_{i \in V_\centernode} q_i.
        \end{equation} 
    \end{proof}
\end{lemma}

As a second baseline algorithm, we test introducing a random time gap between the origination of each misinformation and the delivery of their corresponding prebunks. One way to achieve this time gap with a deterministic policy is to always wait until the misinformation arrives at a node neighboring the central user $c$ before providing user $c$ with a prebunk. We call such policy the \textit{delayed response policy} and provide a description of it in Algorithm \ref{alg:pro}. As was the case for Algorithm \ref{alg:pre}, Algorithm \ref{alg:pro} also guarantees the feasibility condition \ref{eq:opt:feasibility}.

\begin{algorithm}
    \caption{Delayed Response Policy}\label{alg:pro}
    \begin{algorithmic}
       \State For each $k \in \NN$, observe misinformed nodes $I_t^k$ in the local neighborhood of the user.
       \State Choose action
       \begin{equation}
           A_t = \bigg\{P_k \in P\Big | I_t^k \cap \InNeigh(c)\not = \emptyset, \, P_k \not \in \bigcup_{\tau = 1}^{t - 1} A_\tau\bigg\}
       \end{equation}
    \end{algorithmic}
\end{algorithm}

The expected cost of Algorithm \ref{alg:pro} depends on the distributions of the misinformation arrival times $X_k$. While it is difficult to identify the probability distributions of $X_k$ on an arbitrary local network graph $\Graph_c$, In the cases where $q_i(t)$ varies over time, Algorithm \ref{alg:pro} tends to admit better results, as the random time gap between misinformation originations and the arrival of misinformation to the immediate neighborhood of user $c$ results in prebunk deliveries that are more evenly spread out over time, reducing cost $C(\pi)$. We show empirical evidence that supports this effect in the next section.

\section{TEMPORALLY EQUIDISTANT PREBUNKING}
Consider a simpler problem of delivering $\lfloor \epsilon T \rfloor + 1$ prebunks until time $T$, where $\epsilon \in (0,1]$ is a constant, with minimal maximum accumulated cost. This problem can be thought as the original minimax problem \ref{prob:optimization_minimax}, with the restriction that $X_k = T$ for exactly $\lfloor \epsilon T \rfloor$ misinformations and $X_k = 0$. Intuitively, this restriction allows us to search for the best case policy, without being restricted by the constraints that require delivering the prebunks before their associated misinformations. Let $(t_0\dots t_{\lfloor \epsilon T \rfloor})$ denote a non-decreasing sequence of prebunk delivery times, such that we require $t_i \leq T$ for all $i$.

The maximal cost $C(A)$ of delivering these prebunks is
\begin{equation}
    C(A) = \max_{n\in\{0 \dots \lfloor \epsilon T \rfloor\}} \sum_{i = 1}^{n} \beta^{t_n - t_i}.
\end{equation}
And the goal of our optimization is to find,
\begin{subequations} \label{prob:simplified1}
\begin{align}
    \min_{t_1\dots t_{\lfloor \epsilon T \rfloor}} \quad& \max_{n\in\{0 \dots \lfloor \epsilon T \rfloor\}} \sum_{i = 1}^{n} \beta^{t_n - t_j},\\
    \text{s.t.} \quad&  0\leq t_{i-1}\leq t_{i} \leq T, \quad \forall i\in \{1\dots \lfloor \epsilon T \rfloor\}.
\end{align}
\end{subequations}
To find a closed form solution to problem \ref{prob:simplified1}, we relax the condition that $t_1\dots t_{\lfloor \epsilon T \rfloor}$ are integers, and let them take real values. Under this case, Theorem \ref{thetheorem} provides conditions which at least one optimal solution must satisfy.

\begin{theorem}\label{thetheorem}
    Let $t_1\dots t_{\lfloor \epsilon T \rfloor} \in \RR$. Then, there exists a constant $\alpha$, and index $1 < h \leq {\lfloor\epsilon T \rfloor}$ such that there exists an optimal solution to the minimax problem \ref{prob:simplified1} that satisfies
    \begin{subequations} \label{theoremmainresult}
    \begin{align}
        &t_i = 0, \quad \forall n\in \{0\dots h-1 \},\\
        &t_{\lfloor\epsilon T \rfloor} = T,\\
        &\sum_{i = 1}^{n} \beta^{t_n - t_i} = \alpha \geq 1, \quad \forall n\in \{h\dots \lfloor \epsilon T \rfloor\}. \label{thetheorem:maineq}
    \end{align}
    \end{subequations}
    That is, for all $n,m\in \{h\dots \lfloor \epsilon T \rfloor\}$
    \begin{equation}
        \sum_{i = 1}^{n} \beta^{t_n - t_i} = \sum_{i = 1}^{m} \beta^{t_m - t_i}.
    \end{equation}
    \begin{proof}
        Let $c_i = \sum_{i = 1}^{n} \beta^{t_n - t_i}$. It is clear that $t_0 = 0$ and $t_{\lfloor \epsilon T \rfloor} = T$ are optimal choices since all $c_i$ are non-decreasing in $t_0$ and non-decreasing in $t_{\lfloor \epsilon T \rfloor}$.  
        
        Let $t_1\dots t_{\lfloor \epsilon T \rfloor}$ be optimal. Let $h$ be the smallest index for which $t_h > 0$. Also let $t_h\dots t_{\lfloor \epsilon T \rfloor}$ be the minimizer of $\max_{n\in\{h \dots \lfloor \epsilon T \rfloor\}} c_i$. Such an optimal solution is always possible, as costs of $c_i$ for $i<h$ do not depend on $t_h\dots t_{\lfloor \epsilon T \rfloor}$. 
        
        Suppose that $i$ is the smallest number in $\{h, \dots, \lfloor \epsilon T \rfloor\}$ such that $t_i = t_{i+1}$. Then we have $c_i = c_{i+1} - 1 \leq \max_{n\in\{h \dots \lfloor \epsilon T \rfloor\}} c_i - 1$. Thus, there exists a positive $\delta$ for which decreasing $t_i$ by $\delta$ cannot increase $\max_{n\in\{h \dots \lfloor \epsilon T \rfloor\}} c_i$, as $c_i$ is the only value that is decreasing in $t_i$, and all other $c_j$ are either strictly increasing with $t_i$ or do not depend on $t_i$. Thus we can always replace $t_i$ with $t_i-\delta$ to ensure $t_i < t_{i+1}$ without sacrificing optimality. Therefore, there exists an optimal solution in which $t_1,\dots, t_{\lfloor \epsilon T \rfloor} - 1$ exists in the interior of the compact constraint polyhedron,
        \begin{equation}
            P_{\textrm{const}} = \{t_{i-1}\leq t_{i}, \quad \forall i\in \{h\dots \lfloor \epsilon T \rfloor\}.
        \end{equation}
        
        Since we know that $t_{\lfloor \epsilon T \rfloor} = T$ is always optimal, we have only $T-h$ parameters to optimize over. Suppose that $\mathbf{t} = t_h\dots t_{\lfloor \epsilon T \rfloor - 1}$ is in the interior of $P_{\textrm{const}}$, and let $J\subseteq \{h \dots \lfloor \epsilon T \rfloor\}$ denote the set of all indices $j$ such that $c_j = \max_{n\in\{h \dots \lfloor \epsilon T \rfloor\}} c_i$. Clearly, it is possible to reduce this cost, if there exists a vector $\mathbf{u} \in \RR^{T-h}$ such that for all $j\in J$, the $c_j$ is decreasing along $\mathbf{u}$. That is, $(\nabla_{\mathbf{t}} c_j)^\top \mathbf{u} < 0$. We observe that $(\nabla_{\mathbf{t}} c_j) \in \RR^{T-h}$ and the set of vectors $\{\nabla_{\mathbf{t}}c_j|\, j \in J\}$ is linearly independent if and only if $|J|\leq T-h$. Thus, for any $|J|\leq T-h$, it is always possible to find $\mathbf{u}$ such that for all $j \in J$, $(\nabla_{\mathbf{t}} c_j)^\top \mathbf{u} < 0$. Therefore the optimal choice of $t_h\dots t_{\lfloor \epsilon T \rfloor - 1}$ must satisfy $c_j = \max_{n\in\{h \dots \lfloor \epsilon T \rfloor\}} c_i$ for all $j \in \{h \dots \lfloor \epsilon T \rfloor\}$. Defining $\alpha = \max_{n\in\{h \dots \lfloor \epsilon T \rfloor\}}$ completes the proof.
    \end{proof}
\end{theorem}

\begin{corollary}\label{corr}
    Let $t_0\dots t_{\lfloor \epsilon T \rfloor} \in \RR$ be an optimal solution to minimax problem \ref{prob:simplified1}, then $t_{i} - t_{i-1}$ is constant across all $i = h + 1, \dots, \lfloor \epsilon T \rfloor$. Specifically,
    \begin{subequations}
    \begin{align}
        &t_{h} - t_{h-1} = \log_{\beta}(\alpha - 1) -  \log_{\beta}(h), \\
        &t_{i} - t_{i-1} = \log_{\beta}(\alpha - 1) - \log_{\beta}(\alpha).
    \end{align}
    \end{subequations}
    \begin{proof}
    The proof follows directly from applying elementary algebraic manipulations on equation \ref{thetheorem:maineq}.
    \end{proof}
\end{corollary}

Thus, by Corollary \ref{corr}, the optimal policy delivers prebunks evenly across all time steps with equal time gaps in-between. Intuitively, this means that the optimal policy tries to maximize the time gap between each consecutive prebunk delivery. We emphasize that Corollary \ref{corr} only holds exactly when we have the restricted problem \ref{prob:simplified1}, without the additional constraints.

We can approximate this equal time gap policy in the general problem \ref{prob:optimization_minimax}. The goal of the policy 
$\pi$ in this case is to determine which $P_k$ to issue at time $t$. Clearly, it is not optimal to provide a prebunk $P_k$ to user $c$ more than once, since doing so does not change the feasibility of problem and cannot reduce the maximal cost $C(\pi)$. Thus, we can assume without loss of optimality that each $P_k$ is in at most one $A_t \in A$. Hence, in its decision, the policy $\pi$ only needs to take into account the misinformations $M_k$, for which it has not delivered any prebunk $P_k$ yet. Without loss of generality, we assume that this set of misinformations is the finite set $\{M_1, \dots, M_K\}$ as Assumption \ref{assumption:misinfoinit} guarantees that at each time $t$ there are finitely many $M_k$ for which $I_k \neq \emptyset$. 

Suppose that at decision time $t$, the policy can estimate the arrival time of each future misinformation. That is, the policy can estimate $x_k(t) \approx X_k$ as input for all $k$ for which $I_t^k$. Also, let $\sigma:\NN \ra \NN$ represent the permutation in which $x_{\sigma(1}(t), x_{\sigma(2}(t), \dots$ is a non-decreasing sequence. That is, $\sigma$ denotes a non-decreasing sorting of $x_k(t)$.

The cost $C(\pi)$ is only dependent on the timings of prebunk deliveries and not their corresponding misinformation. Thus, for searching an optimal solution, it is useful for us to consider the prebunk delivery times $t_1, \dots, t_K$. Again, without loss of generality we can assume that $t_1, \dots, t_K$ is a non-decreasing sequence, so long as we deliver prebunk $P_{\sigma(i)}$, at time $t_i$. We can make this assumption, because if $t_1, \dots, t_K$ provides a feasible prebunk delivery sequence for any ordering of $\{P_1, \dots, P_K\}$, it will also provide a feasible prebunk delivery sequence for the ordering induced by permutation $\sigma$.

Under these assumptions, we can approximate the equal time gap solution in Corollary \ref{corr} by solving the following linear program,
\begin{subequations}\label{LP}
    \begin{align}
        \max_{u, t_1\dots t_{\lfloor \epsilon T \rfloor}} \quad& u,\\
        \text{s.t.} \quad&  u \leq t_{i+1} - t_i, \quad \forall i \in \{1,\dots,K\},\\
        &t_i \leq x_{\sigma(i)}(t) - 1, \quad \forall i \in \{1,\dots,K\}, \label{LP:mainconst}
    \end{align}
\end{subequations}
where $t_0$ is the time of delivery of the last prebunk before $t$. Note that, $u$ is a slack variable providing a lower bound on $t_{i+1} - t_i$, forcing them to be equal. $u$ also forces $t_i$ to be as widely spaced as possible, so long as they remain less than $x_{\sigma(i)}(t)$. Thus, \ref{LP:mainconst} guarantees that the general problem \ref{prob:optimization_minimax} is feasible under the condition $x_k(t) = X_k$, since we deliver $P_{\sigma(i)}$ at time $t_i$. That is, the feasible solution of the linear program \ref{LP} induces feasible prebunk deliveries to the general problem \ref{prob:optimization_minimax}, so long as we can estimate $X_k$ without error.

In general the policy cannot have an exact knowledge of $X_k$. However, it is often possible to estimate $X_k$. Clearly in this case we lose the feasibility guarantee from before. One method of solving this issue is by defaulting to Algorithm \ref{alg:pro} if there is any misinformation---for which user $c$ has not received prebunks---in the immediate neigborhood of user $c$. 

We present the summary of these methods in Algorithm \ref{alg:lop}. At each time step $t$, Algorithm \ref{alg:lop} first identifies unprebunked misinformations, that is it creates a list of all misinformation indices $k$ such that $I^k_t \neq \emptyset$ and $P_k \not \in A_\tau$ for any $\tau \leq t$. Then if there is a $I^k_t$ that intersects with $\InNeigh(c)$ then it immediately defaults to the decision rule of Algorithm \ref{alg:pro} to guarantee feasibility condition \ref{eq:opt:feasibility}. Otherwise, it solves the Linear program \ref{LP}, and delivers all prebunks that are overdue, that is it provides prebunks $P_{\sigma(i)}$ for all $t_i < t$.

\begin{algorithm}
    \caption{Temporally Equidistant Prebunking}\label{alg:lop}
\textbf{Input} Current time $t$, time of previous prebunk $t_0$
    \begin{algorithmic}
        \State Identify unprebunked misinformations $M_k$. 
        \State ${A_t \gets \bigg\{P_k \in P\Big | I_t^k \cap \InNeigh(c)\not = \emptyset, \, P_k \not \in \bigcup_{\tau = 1}^{t - 1} A_\tau\bigg\}}$
        \If {$A_t \neq \emptyset$}
            \State$t_0 \gets 0$
            \State \Return
        \EndIf
        \State $\forall k \in [K],\quad \hat{X_k} = \textrm{Estimate}(X_k)$ 
        \State Sort into non-decreasing sequence $\hat{X}_{\sigma(1)},\dots,\hat{X}_{\sigma(K)}$
       \State $t_1,\dots,t_K \gets$ solve linear program \ref{LP}.
       \State ${A_t \gets \big\{P_{\sigma(i)} \in P\big| t_{i} \leq t\big\}}$
    \end{algorithmic}
\end{algorithm}

In our experiments, we use a Monte Carlo method and estimate $X_k$ by simulating misinformation propagations in the local network $\Graph_\centernode$, and obtaining a frequentist estimation of how long it takes for the misinformation to reach from any node in the network to the central user $\centernode$. Clearly, this estimation method does not scale to large networks. However, in our tests, for small choices of $m$ the local network around user $\centernode$ remains small enough to permit this approach. It is possible that there are other estimation methods that admit more accurate estimations or scale better to large networks, which might provide better results.

\section{NUMERICAL RESULTS}
\subsection{Algorithm Comparison} \label{experimentsection}
We compare Algorithms \ref{alg:pre}, \ref{alg:pro}, and \ref{alg:lop} by evaluating their performance on simulated misinformation propagations in large scale-free networks. We simulate the misinformation propagation as an SI process on a Chung-Lu model \cite{Chung2002}. 

Constructing a Chung-Lu model with $N$ vertices requires an $N$ dimensional vector $w\in \RR_+^N$ with positive entries. The edge probabilities are then as follows,
\begin{equation}
    p_{ij} = \frac{w_iw_j}{\sum_{i=1}^N w_i}
\end{equation}
An important fact about Chung-Lu networks is that the expected degree of each vertex $i$ is approximately $w_i$, and this approximation tends to be equality for $N\ra \infty$. These networks are abundant in simulating social networks due to having degree distributions that are easy to tune, allowing them to emulate a wide range of different network topologies.

To model a realistic proxy of a social network we require the Chung-Lu network to be scale-free, meaning that the degree distributions, weighted by $p_{ij}$, need to follow a power-law. Thus we choose
\begin{equation}
    w_i = C\, i^{-\frac{1}{\gamma-1}},
\end{equation}
where $C$ and $\gamma$ are tunable parameters. By Lemma 3.0.1 from Fasino et al. \cite{FasinoDario2021Glsn}, this choice for $w_i$ guarantees the resulting Chung-Lu model is scale-free.

We selected the parameters $C$ and $\gamma$ by tuning them so that at each step, the resulting SI propagation most closely resembles the daily message propagation in a real social networking platform. To do so, we chose the Twitter WICO dataset \cite{wico2021}, which is a collection of timestamped propagations containing tweets/re-tweets, in two misinformation and one non-misinformation categories. We then tuned the the parameters $C$ and $\gamma$ so that the number of infected users in the simulated SI cascade on the resulting Chung-Lu model matches the misinformation propagation in the WICO dataset. Fig. \ref{fig:small_network} shows a small example of this Chung-Lu network consisting of $100$ users.

\begin{figure}[t]
    \centering
    \includegraphics[width = \linewidth]{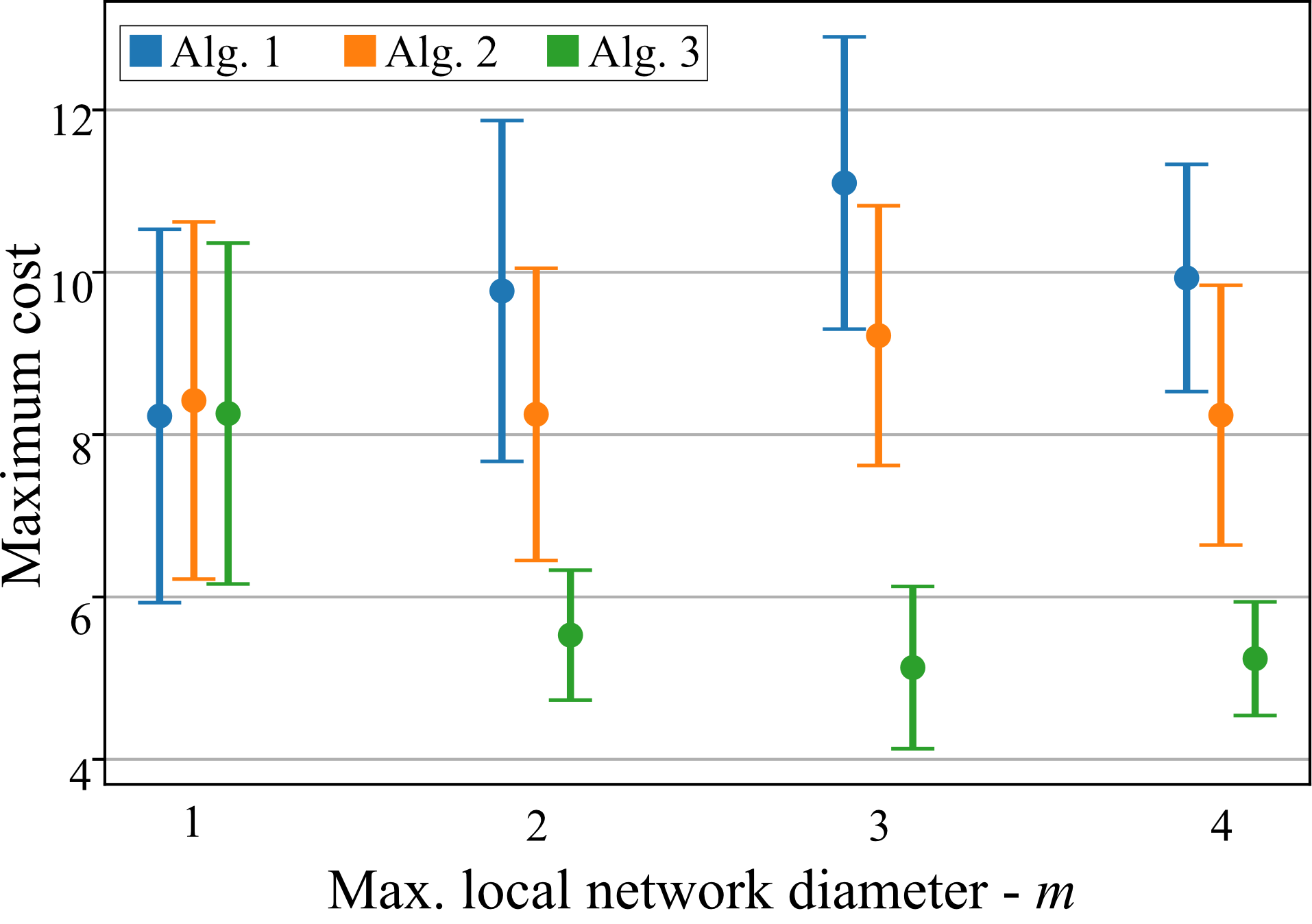}
    \caption{Comparison of three different prebunking policies induced by Algorithms \ref{alg:pre}, \ref{alg:pro}, and \ref{alg:lop} across different spatial horizon parameters $m$, denoting the maximum local network diameter. The vertical axis represents $C(A)$, the maximum accumulated cost.}
    \label{fig:maintest}
\end{figure}

For comparison of Algorithms \ref{alg:pre}, \ref{alg:pro}, and \ref{alg:lop}, we construct a larger Chung-Lu model consisting of $2000$ users. We also prune the edges with $p_{ij}$ less than $5\times10^{-4}$. This pruning speeds up computation and allows us to have non-trivial local neighborhoods in the network.  We then simulate sample misinformation propagations from time $t \in [0,100]$ by following the propagation law from Assumption \ref{assumption:model}. At each time step, we let each user in the network initiate a new misinformation cascade with probability $q_i(t) = 5\times 10^{-5}$. A slight deviation we take from Assumption \ref{assumption:misinfoinit} is that we limit the maximum number of allowed misinformation on the network to $50$ for computational restrictions. This means that the actual misinformation generation probabilities $q_i(t)$ effectively decrease exponentially over time. 

We test each algorithm by first choosing  random user to guard against misinformation, which we call $c$ as before. We then construct a local neighborhood of radius $m$ around $c$ by finding the subgraph containing all users that are at most $m$ graph distance away from $c$. Fig. \ref{fig:maintest} shows the distributions of the accumulated maximal cost $C(A)$ across four different choices of $m$. For each $m$ we repeat the experiment $10$ times and plot the mean and the deviations of $C(A)$ across these runs.

\begin{figure}[t]
    \centering
    \includegraphics[width = \linewidth]{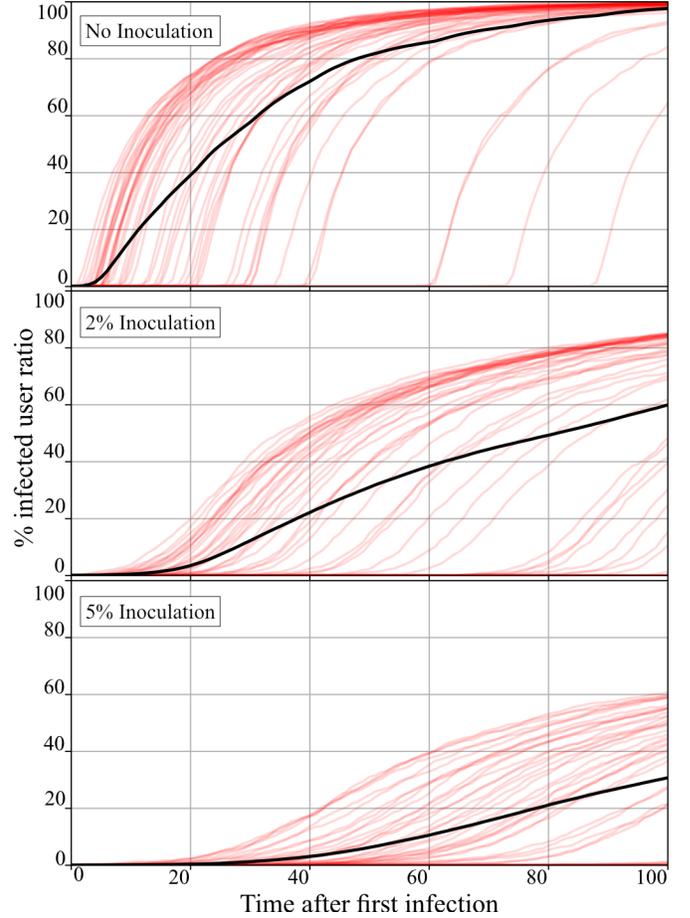}
    \caption{Comparison of three different inoculation ratios in terms of percent ratio of misinformed users across time. Each red line shows a misinformation cascade starting from the first time the misinformation appears in the network. The bold black lines show the average ratio of misinformed users.}
    \label{fig:cascades}
\end{figure}

Fig. \ref{fig:maintest} shows that Algorithm \ref{alg:lop} clearly achieves prebunking with less cost than the baselines, with its average cost being approximately half of the other two algorithms for $m \geq 2$. Note that all three algorithms achieve the same cost for $m = 1$ as the feasibility restrictions force all three algorithms to induce the same policy in this case. That is, their behavior are identical for $m = 1$. Also note that since the diameter of the original network is $4$, we restrict $m$ to be less than $4$.

\subsection{Effects of Prebunking on the Misinformation Propagation}
So far we have discussed policies to guard a single user $c$ against misinformation by issuing them prebunks before their corresponding misinformation content. In a social networking platform, however, the goal is often to guard as many users as possible against misinformation. In this paper we mainly focus on optimizing the prebunking delivery to a single user since protecting singular users $c$ extends naturally to protect the entire network, as it is possible to significantly reduce the propagation of misinformation by applying these prebunking algorithms to a selected group of users, inoculating them against misinformation and preventing them from spreading misinformation to users they influence.

We test the effect of Algorithm \ref{alg:lop} on the misinformation propagation by simulating a misinformation cascade on a Chung-Lu network with $2000$ users, using the same parameters as in section \ref{experimentsection}. However, instead of providing prebunks to a single user, we provide prebunks to a small subset of users with maximal expected degrees. We show the resulting misinformation propagation curves in Fig. \ref{fig:cascades}. It is evident that applying Algorithm \ref{alg:lop} to deliver prebunks to a small number of nodes significantly delays the propagation of misinformation, at least for the scale-free networks with which we are experimenting. This is not surprising in and of itself, as it is consistent with the existing literature which demonstrates that the shortest path lengths between nodes in a scale-free network is often sensitive to the removal of the highly connected nodes \cite{guillaume2005comparison, CRUCITTI2003622}. In addition, prebunking reduces the variance between different misinformation cascades, making their propagation more predictable.

\section{CONCLUSION}
We define the problem of optimally delivering prebunks to a user as a minimax optimization problem, and under SI propagation assumptions propose algorithms that guarantee feasibility. We demonstrate that our theoretically backed approach Algorithm \ref{alg:lop} also yields better results than the other two baselines in empirical analysis using simulated misinformation propagations on Chung-Lu models. Algorithm \ref{alg:lop} is also often computationally feasible to solve, as at each time, it relies on solving a linear program that is computationally inexpensive. 

Our results, however, suffer from the limitations we impose on the network structure. Real-world misinformation propagation often deviates significantly from the SI model predictions. Our models also only provide feasibility guarantees under the discrete-time epidemic propagation assumptions. We also focus solely on delivering optimal prebunks to each user one by one, which is different from optimizing prebunk deliveries on the entire network. For future work, we plan to extend our problem and results to optimizing misinformation deliveries across the entire social network.

\printbibliography
\end{document}